\RequirePackage[2020-02-02]{latexrelease}
\documentclass[pra, amsmath, reprint, amssymb, aps, superscriptaddress, twocolumn]{revtex4}
\usepackage{graphicx}
\usepackage{amsmath, braket, amsfonts}

\usepackage{multirow}
\usepackage{mathrsfs,amsmath} 
\usepackage{fixmath}
\usepackage{graphicx}
\usepackage[colorlinks,linkcolor=blue,anchorcolor=red,citecolor=green]{hyperref}
\usepackage{placeins}
\usepackage{cancel}

\begin{document}


\title{Proposal to detect emergent gauge field and its Meissner effect in spin liquids using NV Centers}

\author{Patrick A. Lee}
\affiliation{Department of Physics, Massachusetts Institute of Technology, Cambridge Massachusetts 02139, USA}

\author{Sid Morampudi}
\affiliation{Department of Physics, Massachusetts Institute of Technology, Cambridge Massachusetts 02139, USA}

\date{\today}

\begin{abstract}
    We show that NV centers could provide distinct signatures of emergent gauge fields in certain phase transitions in spin liquids. We consider the relaxation rate $1/T_1$ of a diamond NV center placed a distance $z_0$ above a crystal consisting of two dimensional layers which hosts a U(1) quantum spin liquid with  spinon Fermi surface. We show that during a spinon pairing transition from a U(1) to a Z2 spin liquid, the relaxation rate exhibits a rapid fall just below the pairing transition. The drop is much faster than what can be accounted for by the opening of the energy gap. Instead, the rapid fall with a unique $z_0$ dependence of its height and width, is the signature of the Meissner effect of the gauge field. We identify the organic ET compound which has a phase transition at 6K as a candidate for the experimental observation of this phenomenon.
\end{abstract}



\maketitle

\section{Introduction}

Quantum spin liquid represents an exotic state of matter where a Mott insulator or a spin system with antiferromagnetic coupling fails to order magnetically at zero temperature due to quantum fluctuations~\cite{Savary2016,Zhou2017}.  It has been under intense studies both theoretically and experimentally because it is a prime example of the notion of emergence: novel particles and field emerge in the low energy description which are absent in the original Hamiltonian. These emergent particles make their appearance most naturally in a parton construction, where the spin operator is represented by spinful fermions or bosons or where an electron in a Hubbard model is represented by fermions which carry spin (called spinons) and a boson which carry charge (called chargon). This procedure introduces extra degrees of freedom and gauge fields are introduced which project the system back to the physical degrees of freedom. The magic is that in certain cases, these artificial degrees of freedom which are initially confined by the strong gauge fluctuations become emergent deconfined particles coupled with emergent gauge fields at low energy and long distances. This mean-field like picture does not usually have a well controlled expansion parameter and its validity has been a subject of debate. However, the notion of emergent particles and fields receive a great boost by the discovery of exactly soluble models, such as the Kitaev model on the honeycomb lattice~\cite{kitaev2006}. 
 Starting with a spin model, it was clearly shown that the low energy excitations are fermionic spinons (gapped or gapless) and gapped Z2 gauge fluxes called visons.
This fits perfectly the mean-field description of fermionic spinons coupled to Z2 gauge fields~\cite{burnell2011}.

Experimentally there has been a decades long ongoing search for materials which exhibit spin liquid behavior. While there is a long list of magnetic systems that does not order, it is not so easy to prove that a given material is a genuine spin liquid. Part of the problem is the effect of disorder which can lead to local moment formation.  This can drive the system to a random singlet phase which does not exhibit long range order, but does not qualify as a spin liquid because it does not exhibit emergent properties~\cite{kimchi2018valence,kimchi2018}. A promising approach is to focus on examples which exhibit special properties which are readily measurable. One recent example is $\alpha-\rm{RuCl_3}$, a honeycomb structure which contains the Kitaev interaction and may exhibit Majorana edge excitations in the presence of a magnetic field. However, the experimental situation is still evolving at this point in time~\cite{czajka2021}. 

Another system which has a well defined signature is the case when the spinons form a Fermi surface. In this case certain properties such as specific heat, spin susceptibility and thermal transport will behave like a metal. In this connection, a class of materials that has been studied for 20 years is the family of organic salts which form approximate triangular lattice~\cite{Zhou2017}. These systems are believed to be described by a Mott-Hubbard model with a tunable hopping parameter. Two examples, called the ET and dmit salts as shorthand notation, are believed to be in close proximity to the insulator side of the Mott transition where a spin liquid state may be expected~\cite{lee2005,Motrunich2006}. Experiments show a linear T specific heat which for many years has been taken as a signature of a spin liquid with spinon Fermi surface. In the case of dmit the thermal conductivity is also linear in $T$ as expected for a neutral Fermi sea, but this data has been challenged~\cite{bourgeois2019thermal,ni2019absence}. On the other hand, ET shows gap like behavior in the thermal conductivity and a recent electron spin resonance experiment shows conclusively that this system has a spin gap of about 15K~\cite{miksch2021gapped}. Intriguingly this system is known to have a phase transition at 6K and a spin gap appears to develop below this transition. We will return to this point later. We also mention that there have been  recent data in monolayer $\rm{1T-TaSe_2}$ which have been interpreted in terms of a spinon Fermi surface~\cite{ruan2021evidence}. Another system that has been interpreted as hosting a spinon Fermi surface that undergoes a transition to a gapped spin liquid is the distorted Kagome system $\rm ZnCu_3(OH)_6SO_4$ ~\cite{Li, Zorko}

While there have been claims and counter-claims concerning spinon sightings, up to now there has been no experimental signs of the emergent gauge field at all.
There have been a few proposals to search for gauge fields. In the case of spinon Fermi surface, the gauge field is expected to be U(1), but overdamped due to coupling to spinons. One proposal is to use an external magnetic field to generate a gauge magnetic field and search for quantum oscillations in the spinon response~\cite{motrunich2006orbital}. A related proposal is to search for thermal Hall conductivity as a response to the gauge magnetic field~\cite{katsura2010theory}. In certain special cases, it has been suggested that gauge field excitations may be seen in neutron scattering~\cite{lee2013proposal}. Unfortunately to the extent that experimental searches have been attempted, there have not been any successful report so far. 

Yet another idea is to monitor the fate of the U(1) gauge field in a system which undergoes a phase transition from a U(1) to a Z2 spin liquid. Within the parton construction, in order for this phase transition to occur  without unit cell doubling, the natural way to describe such a transition is by spinon pairing. Just as in conventional superconductors, the gauge field is gapped by the Higgs mechanism, a Meissner state is formed, and the gauge degree of freedom is broken from U(1) to Z2. In clean superconductors, the onset of the Meissner effect causes a dramatic drop (called the "rapid fall") in the transverse sound attenuation just below $T_c$.\cite{leibowitz} This is because transverse sound couples to electromagnetic field, which is rapidly screened by the onset of superfluid density. It was shown that even though the spinons are charge neutral, they couple to sound waves in the same way as electrons, and a similar phenomenon of "rapid fall" is predicted to occur if a spinon Fermi surface undergoes a transition to a gapped Z2 spin liquid~\cite{zhou2011spinon}. The ET system with the 6K transition mentioned above is a prime candidate to search for this phenomenon. Needless to say, the detection of the Meissner effect associated with the gauge field would constitute a smoking-gun signature. Unfortunately, transverse sound experiments require a relatively thick crystal and this experiment has not been attempted up to now, to our knowledge. 

The goal of this paper is to propose yet another path to measure the onset of Meissner effect of the emergent gauge field. We are inspired by the recent  advances in using NV centers in diamond as a highly sensitive tools to detect magnetic fluctuations a short distance outside the sample. Proposals have been made to detect the magnetic field fluctuations generated by transverse current fluctuations in metals `\cite{Casola2018,Agarwal} and in superconductors ~\cite{dolgirev2022characterizing} and such fluctuations have been successfully detected in metals~\cite{Ariyaratne2018}. Proposals have also been made to detect the magnetic fluctuations outside a spin liquid. There are two mechanisms. The spins in the sample has a dipolar coupling with the effective spin of the NV center. However, this coupling is weak and the signal very small~\cite{Chatterjee2019}. A second path is to consider a spin liquid with a spinon Fermi surface. It was recently shown that the fluctuations in the spinon current density will generate magnetic fluctuations just like a metal, albeit with a reduction factor~\cite{Khoo2022}. This reduction factor depends on the diamagnetic susceptibility of the spinon. In this paper we extend this work to the onset of spinon pairing. We show that the spinon Meissner effect gives rise to a rapid fall of the relaxation rate of the NV center. The rate falls within a very narrow region of temperature just below $T_c$, giving a detectable signal in a realistic system. In contrast, this rapid fall does not occur in the usual metal to superconductor transition. In that case, the relaxation rate decreases gradually due to the continuous opening of the gap~\cite{dolgirev2022characterizing}.

\section{relaxation of the NV center due to transverse conductivity}\label{Sec.NV}

  Current fluctuations in a metal generate magnetic field at the NV center located at a distance $z_0$ above the metal surface  due to Biot-Savart law. The current fluctuation is proportional to the imaginary part of the  transverse conductivity $\sigma _{\perp} (q,\omega)$. 
\
Therefore the quasi-static transverse conductivity can be probed by measuring the $T_1$-time of a single spin (NV center) placed above the sample
~\cite{Casola2018,Agarwal,Ariyaratne2018,Khoo2021}. The $1/T_1$ is  proportional to the imaginary part of the magnetic field autocorrelation function (magnetic noise) at the NV center location, ${\rm Im}\chi _{B_\mu B_\nu} (z, \omega)$, and at the frequency $\omega$ given by the energy splitting of the NV center~\cite{Casola2018,Langsjoen2012}. 
\
We quote the following formula~\cite{Casola2018,Langsjoen2012}:

\begin{align}\label{T1Eq}
\frac{1}{T_1}=\frac{\mu_B^2}{2\hbar} \coth \Bigl(\frac{\beta \hbar \omega }{2}\Bigr) {\rm Im} \chi _{B_z B_z}  (z, \omega),
\end{align}
For a circular Fermi surface of a two dimensional layer, we have:
\begin{align}\label{Imchi1}
{\rm Im} \chi _{B_z B_z}  (z, \omega)
&= \frac{\mu _0 ^2 \omega}{8 
\pi}\int q d q e^{-2qz} {\rm Re}~\sigma _{\perp} (q,\omega).
\end{align}

The  factor $\exp(-2qz)$ indicates that we are probing mainly $q$ of order $1/2z$. In a clean metal ${\rm Re} \sigma_{\perp} (q,\omega) = (e^2/h) p_F/q $
 describes Landau damping due to particle hole excitations. Putting this into Eq. \ref{Imchi1} we find
\begin{align}\label{Imchiz}
{\rm Im} \chi _{B_z B_z} 
&\simeq  \frac{e^2 \mu _0 ^2}{16 \pi h} \frac{\omega}{z}{(2S+1)p_F},
\end{align}
where $p_F$ is the Fermi momentum and we can take $S=1/2$.

\section{Spinon Fermi surface}




In a recent paper Khoo et al ~\cite{Khoo2022} have shown that a similar relaxation rate occurs if the metal is replaced by a quantum spin liquid with a spinon Fermi surface. The difference is that  Eq \ref{Imchiz} is now now mutiplied by a reduction factor $F_0$

\begin{align}\label{F}
F_0= \left( \frac{\chi_c}{\chi_s+\chi_{a}+\chi_c} \right)^2
\end{align}

The physics behind this factor is the following. Even though the spinons are charge neutral, they couple to an emergent gauge field $a$ which in turn is coupled linearly to the physical electromagnetic gauge field $A$. This is because in order to enforce the constraint of no double occupation, the spinon current must be balanced by a chargon current which does carry physical charge. This phenomenon results in the Ioffe-Larkin rule ~\cite{Ioffe1989} and the factor  $\alpha = \frac{\chi_c}{\chi_s+\chi_{a}+\chi_c}$ can be viewed as the ratio of the physical current to the spinon current. Since the conductivity involves two factors of current, the correction $F_0$ to conductivity and $1/T_1$ appears as the square of this ratio. 
In deriving these results we assume for following form for the spinon transverse conductivity 
\begin{align}\label{sigmap}
\sigma^{\rm spinon}_{\perp}
(q,\omega) \approx \frac{e^2}{h} \frac{p_{\rm F}}{q}+ \chi_s \frac{q^2}{i \omega}.
\end{align}

In Eq. \ref{sigmap} the first term is due to particle-hole excitation near the Fermi sea. It is valid for $q>1/l_{\rm mfp}$ where $l_{\rm mfp}$ is the meann free path due to disorder scattering. Hence the $q$ integral in Eq \ref{Imchi1} should be cutoff by $1/l_{\rm mfp}$.
  In the second term, $\chi_s$ is the Landau diamagnetic susceptibility. For a non-relativistic parabolic dispersion for the spinons one obtains $\chi_s=g_s/(24\pi m_s)$ ,which is the standard Landau diamagnetic constant with $g_s=2$ accounting for the spin degeneracy.

The chargon is gapped and within a relativistic boson model of the chargon dispersion, one obtains $\chi_c=1/(24\pi m_c)$, where $m_c= \Delta_c/v_c^2$, is the effective mass of the chargons with a speed $v_c$ and a gap $\Delta_c$, as shown in Ref.~\cite{lee2005,dai2020modeling}. Note that we will be interested in spin liquid that are stabilized by being proximate to that Mott transition \cite{lee2005,Motrunich2006} where $\Delta_c$ is small and $\chi_c$ is large. Finally $\chi_a$ is the coefficient of the $b^2$ term in the effective Lagrangian for the gauge field, where b is the gauge magnetic field. It comes from integrating out energy on the upper cut-off scale. This term was not in the Ioffe-Larkin formula but introduced in \cite{Khoo2022} as an extension of Ioffe-Larkin. For our problem it is expected to be small compared with $\chi_c$ which is enhanced by the small gap $\Delta_c$ and can be ignored. If $\chi_c$ also dominates over $\chi_s$ the reduction factor $F$ can be close to unity. This predicts metallic like contributions to $1/T_1$ in an insulator and can be considered a signature of spinon Fermi surface if observed. The order of magnitude of the resulting $1/T_1$ was estimated in \cite{Khoo2022}. 

%
%

\

\section{Rapid fall of the relaxation rate below the spinon pairing transition}\label{Sec.rapid}

We now turn our attention to the case when the spinon Fermi surface undergoes a pairing transition. In the case of a full pairing gap, this is the route  to go from U(1) to Z2 spin liquid wihin the gauge theory approach. First we mention that the case of conventional metals undergoing a superconducting transition has been treated\cite{dolgirev2022characterizing}.  In that case only the real part of $\sigma_\perp$ which represents dissipation enter the expression to $1/T_1$. This corresponds to the  first term in Eq. \ref{sigmap} representing particle-hole excitations which become gapped in the superconductor. Since superconductivity is a continuous transition, the gap opening affects $1/T_1$ very gradually just below $T_c$. The Meissner effect is reflected by a strong enhancement of $\chi_s$ in the second term in Eq. \ref{sigmap}, but since it is purely imaginary, it does not affect $1/T_1$ in the superconductor. The spinon pairing case has an important difference. In this case the diamagnetic susceptibility $\chi_s$ in the second term in Eq. \ref{sigmap} enters via the factor $F_0$. Instead of Landau diamagnetism, Meissner effect of the gauge field sets in. In a two fluid model, just below $T_c$ we replace $\chi_s q^2$ by $\chi_s q^2 + \rho_s(T)/m_s$ where $\rho_s$ is the spinon superfluid density which takes the form $\rho_s(T) = \rho_0 (T_c-T)/T_c$ in meane field theory.  Here $\rho_0 = 1/a^2$ is the spinon density and $a^2$ is the unit cell area. We will be interested in temperatures very close to and just below $T_c$ because we will see that the "rapid fall" of $1/T_1$ happens there. Hence the real part of Eq.\ref{sigmap} is assumed to stay the same as in the metal and we ignore the small change in the normal density in the two fluid model. The important change is in the second term in Eq.\ref{sigmap}. Since $q \approx 1/2z$ is small on the atomic scale, the replacement of $\chi_s q^2$ by $\chi_s q^2 + \rho_s(T)/m_s$ has a dramatic effect on the factor $F$ which changes rapidly immediately below $T_c$. It is given by 
 
\begin{align} \label{F1}
    F_1= \frac {F_0} {1+ \gamma_1 (z/a)^2(T_c-T)/T_c}
\end{align}
where $\alpha=\chi_c/(\chi_c+\chi_0 +\chi_s)$ as defined earlier and $\gamma_1 = 4 \alpha/(m_s\chi_c)$ As discussed earlier we expect $\chi_c$ to dominate $\chi_0$, so $\alpha$ is close to unity. The parameter $\gamma_1$ is not so well known, but may be of order unity in practice. Combining Eqs. \ref{T1Eq},\ref{Imchiz},\ref{F1} and taking the limit $T > \omega$, we find for single layer

%
\begin{align} \label{T11}
1/T_1 = \frac{\mu_B^2}{\hbar} \frac{e^2 \mu _0 ^2 p_F}{8\pi h}(T/z)F_1
\end{align}

Note that $F_1$ drops rapidly below $T_c$. It drops to 1/4 of the value above $Tc$ when $(T_c-T)/T_c = (a/z)^2/\gamma_1$. With increasing $z$ the value of $1/T_1$ decreases while the width below the transition rapidly decreases as  $(a/z)^2$. This is the phenomenon of "rapid fall" and has its origin in the Meissner effect of the emergent gauge field.

\begin{figure}[htb]
\begin{center}
\includegraphics[width=3.5in]{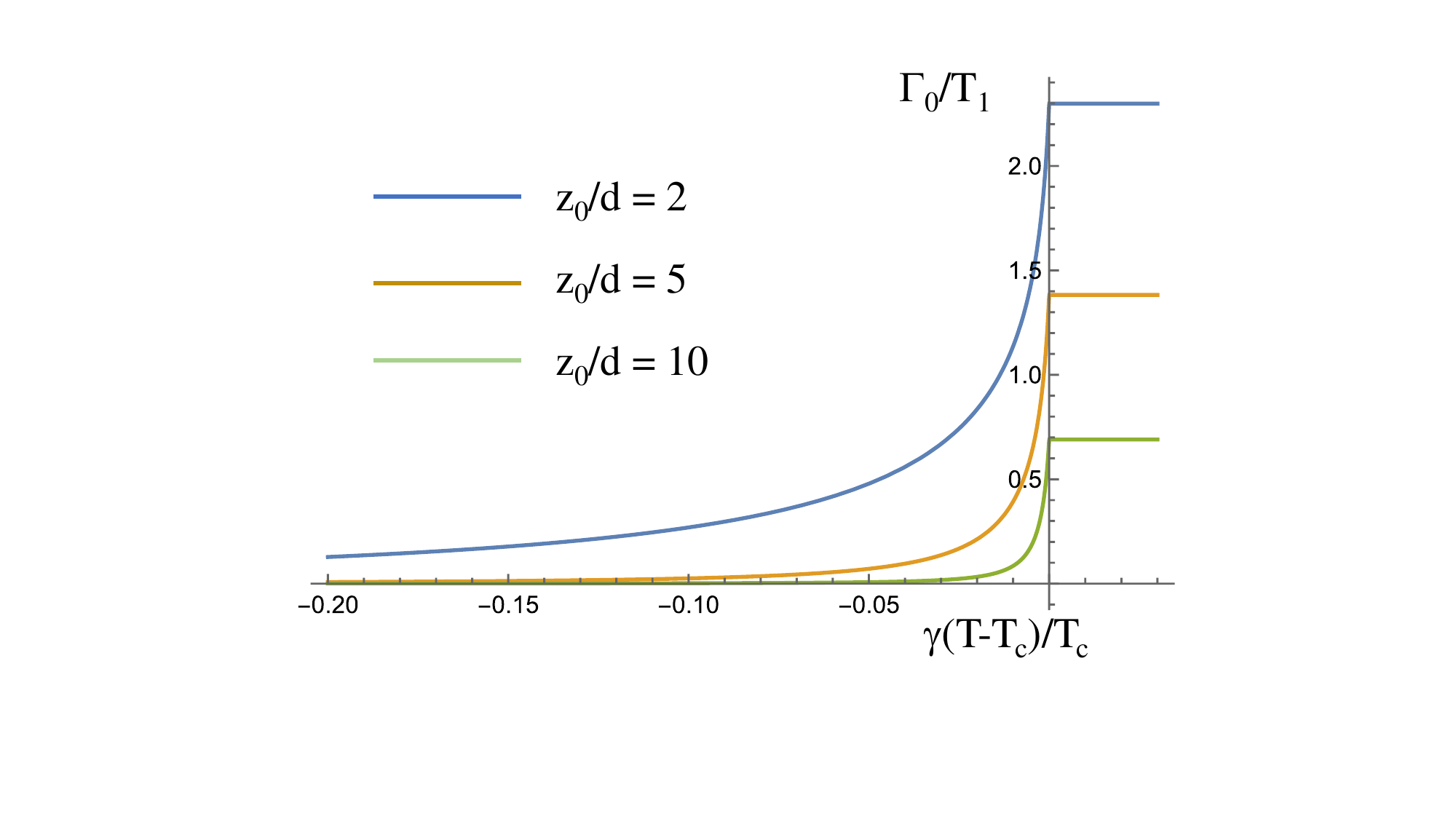}
\caption{The relaxation rate $1/T_1$ for a NV center placed at a distance $z_0$ above a crystal consisting of layers that are separated by a distance $d$. Plots are shown for several values of $z_0/d$. Note the rapid fall of the signal below the onset of spinon pairing with a width which decreases as $(d/z)^2$. The vertical scale is measured in units of $\Gamma_0$ which is $1/T_1$ for a single layer without pairing at a distance $d$ and $\gamma$ is a number of order unity (see text). The scattering mean free path $\rm mfp$ is set to be $20d$}
\label{Fig: T1}
\end{center}
\end{figure}

The $T_1$ time for a single layer of the organic spin liquid candidate was estimated in the spinon Fermi surface state \cite{Khoo2022} and found to be of order 100 msec at 10K and at a distance of 1 nm. This makes detection difficult. The situation improves if we consider a stack of multi-layers with layer spacing $d$, as in the case for a single crystal and place the NV center at a distance $z_0$ above the surface. We can sum over contributions from the top N layers where N is set to $l_{\rm mfp}/d$ to reflect the cutoff due to disorder. Assuming $z_0 > d$ and under the condition $z_0 < l_{\rm mfp}$ we can replace the sum with an integral, resulting in

\begin{align} \label{T12}
1/T_1 = \Gamma_0 \int_{z_0/d}^{l_{\rm mfp}/d} \,dx \, \frac{1}{x(1+[\gamma (T_c-T)/T_c] x^2)}
\end{align}
where $\gamma= \gamma_1(d/a)^2$ is a number of order unity and the overalll factor $\Gamma_0$ is defined as $1/T_1$for a single layer at a distance d just above $T_c$ as given by Eq \ref{T11}. Note that $\Gamma_0 \propto T/z$, its absolute value can easily be obtained using the estimate in Ref. ~\cite{Khoo2022} which was made for $T=10K$ and $z=1 nm$.
The indefinite integral in Eq. \ref{T12} , $\int dx \, 1/[x(1+a x^2)^2]  $ can be evaluated to give $\ln x +[1/(1+a x^2) -  \ln(1+a x^2)-1]/2$. Examples for $l_{\rm mfp}/d=20$ and several values of $z_0/d$ are plotted in Fig. 1. Note that the multilayer case has the advantage that the size of the signal decreases slowly (logarithmically) with increasing $z_0$, (as opposed to $1/z$ for a single layer)  making its detection more feasible for realistic $z_0$ which is often of order 10 nm. The important point is that the width is very narrow below $T_c$ and scales as $(d/z_0)^2$. This rapid fall phenomenon is the signature of the Meissner effect of the gauge field.

\section{Conclusion}

We have shown that the onset of spinon pairing in a U(1) spin liquid leads to a sudden drop in the relaxtion rate $1/T_1$ of a NV center. This drop results from the Meissner effect of the emergent gauge field. Its detection will be a strong confirmation of the notion of emergent spinons and gauge fields. We identify the organic ET compound as a promising candidate where this effect may be measurable.

\
\
Acknowledgements.  We thank Shubhayu Chatterjee, Ruolan Xue and Kehang Zhu for discussions on the experimental aspects of NV centers.  P.L. acknowledges the support by DOE office of Basic Sciences Grant No. DE-FG02-03ER46076.
\bibliographystyle{apsrev4-1}
\bibliography{bibfile}

\clearpage

%
%
%

\end{document}